\documentclass[a4paper]{paper} 
\usepackage[margin=2.5cm]{geometry}
\usepackage{lipsum}
\usepackage{xcolor}
\usepackage{booktabs}
\sectionfont{\large\sf\bfseries\color{black!70!blue}} 
\usepackage{graphicx} 
\usepackage{authblk}

\usepackage{braket}
\usepackage{physics}
\usepackage{multirow}
\usepackage{siunitx}
\newtheorem{theorem}{Theorem}
\newtheorem{remark}[theorem]{Remark}
\newcommand\Tstrut{\rule{0pt}{2.6ex}}         
\newcommand\Bstrut{\rule[-0.9ex]{0pt}{0pt}}   
\usepackage{subcaption}
\usepackage{algorithm}
\usepackage[algo2e]{algorithm2e} 
\usepackage{hyperref}
\usepackage{cleveref}
\usepackage{biblatex}
\bibliography{Ref.bib}

\begin{document}
\title{\centering Stressing Out Modern Quantum Hardware: Performance Evaluation and Execution Insights}
\author[1,2]{Aliza U.~Siddiqui}
\affil[1]{\small Department of Electrical, Computer, and Energy Engineering, University of Colorado Boulder,
Boulder, Colorado 80309, USA }
\affil[2]{\small Department of Atomic and Laser Physics, University of Oxford, Oxford OX1 2JD}
\author[2]{Kaitlin ~Gili}
\author[2]{Chris Ballance}

\maketitle

\begin{abstract}
    Quantum hardware is progressing at a rapid pace and, alongside this progression, it is vital to challenge the capabilities of these machines using functionally complex algorithms. Doing so provides direct insights into the current capabilities of modern quantum hardware and where its breaking points lie. Stress testing is a technique used to evaluate a system by giving it a computational load beyond its specified thresholds and identifying the capacity under which it fails. We conduct a qualitative and quantitative evaluation of Quantinuum's H1 ion-trap device using a \emph{stress test} based protocol. Specifically, we utilize the quantum machine learning (QML) algorithm, the Quantum Neuron Born Machine, as the computationally intensive load for the device. Then, we linearly scale the number of repeat-until-success sub-routines within the algorithm to determine the load under which the hardware fails and where the failure occurred within the quantum stack. Using this proposed method, we (a) assess the hardware’s capacity to manage a computationally intensive QML algorithm and (b) evaluate the hardware performance as the functional complexity of the algorithm is scaled. Alongside the quantitative performance results, we provide a qualitative discussion and resource estimation based on the insights obtained from conducting the stress test with the QNBM. 
\end{abstract}

\section{Introduction}

Quantum computing technologies have progressed rapidly over the past few years, including recent improvements in quantum volume and the implementation of more advanced features inspired by quantum theory \cite{moses2023race, Baldwin2022reexaminingquantum,decross2022qubitreuse,Pino2021,QuantinuumHardware}. Alongside this progression comes the responsibility to challenge these machines with functionally complex algorithms that align with the new hardware capabilities. For example, NISQ algorithms \cite{Preskill2018quantumcomputingin, NISQ2022}, while primarily developed to leverage limited quantum resources to perform classically challenging tasks, have also served as useful evaluation tools for present day quantum hardware \cite{osti_1658010, PhysRevA.99.062323, benny_qcbm, wang_bench, benchmark_sim, obst2023comparing, mesman2022qpack}. As we now make the transition from NISQ devices to fault tolerance, it remains useful to \emph{stress test} - identify the non-obvious breaking points of - the hardware when running algorithms containing these more complex features at scale.   In doing so, we assess the hardware's current capabilities, identify the specific road-blocks for scale, and obtain specific insights for future improvement. 

Recently, a quantum generative algorithm known as the \textit{Quantum Neuron Born Machine (QNBM)}, was developed \cite{GiliOG, gili2023generative} as a potential application for advanced NISQ devices. The QNBM is a high-depth, high-width circuit model that requires \textit{combined} intensive functionalities such as mid-circuit measurements, repeat-until-success (RUS) routines, classical control, and large degrees of entanglement. These features make the QNBM a very demanding algorithm for quantum hardware, and in fact, only in the  latest versions of a few current devices contain the necessary operations for execution. As such, the QNBM is a strong candidate to stress test advanced NISQ hardware.

In this work, we introduce an interpretation of a stress test (used in software engineering)\cite{StressTesting} with the QNBM algorithm which can be tuned in functional complexity by scaling the algorithm's number of RUS routines. With this proposed method, we choose to conduct a quantitative and qualitative performance evaluation of the  Quantinuum H1-1 trapped ion device \cite{QuantinuumHardware} interfacing with the $Q-\#$ programming language, as this system contains support for the advanced capabilities aforementioned. More specifically, we (a) assess the hardware's capacity to manage a computationally intensive Quantum Machine Learning (QML) algorithm and (b) evaluate the hardware performance as the functional complexity of the algorithm is scaled. Alongside the quantitative performance results, we provide a qualitative discussion based on the insights obtained from conducting the stress test with the QNBM - i.e. we put forth the specific programming and hardware-induced road blocks that appear when scaling the functional complexity of the algorithm. We want to emphasize that this interpretation of the stress test is hardware-agnostic, meaning that it can be utilized on other quantum devices such as IBMQ and IonQ as an evaluation protocol \footnote{At the time of this work, popular quantum devices such as IBMQ and IonQ did not support functionalities such as classical control and mid-circuit measurements}.

The remainder of the paper is organized as follows: we begin with the details of the proposed methodology for evaluation in Sec. \ref{summary}. In Sec. \ref{quantitative}, we provide a brief summary of the selected hardware for stress testing and an overview of the specific details chosen for the evaluation. We then 
execute the methodology on Quantinuum's hardware, answering the following questions: \textit{``how does the hardware perform when executing an algorithm of this functional complexity?"} and \textit{``how does the hardware perform when we scale the algorithm functional complexity
i.e. add more RUS routines or model output neurons?"}. We then discuss the insights learned from our assessment and present a discussion on future ways to improve the software-hardware integration in Sec. \ref{qualitative}. Overall, we hope the positive quantitative results presented in this work are encouraging for algorithm designers who are looking to utilize near term hardware, and simultaneously, we hope the programming and hardware breaking points discussed here encourage architecture researchers to further enhance the quantum stack with these insights in mind. 

\section{Proposed Evaluation Method}\label{summary}

In this section, we introduce our proposed evaluation method for advanced NISQ hardware. We first provide an overview of the QNBM algorithm prior to describing how it can be used in a stress test interpretation. 

\subsection{QNBM Algorithm}

\begin{figure}[b!]
\begin{center}
    \includegraphics[scale=0.4]{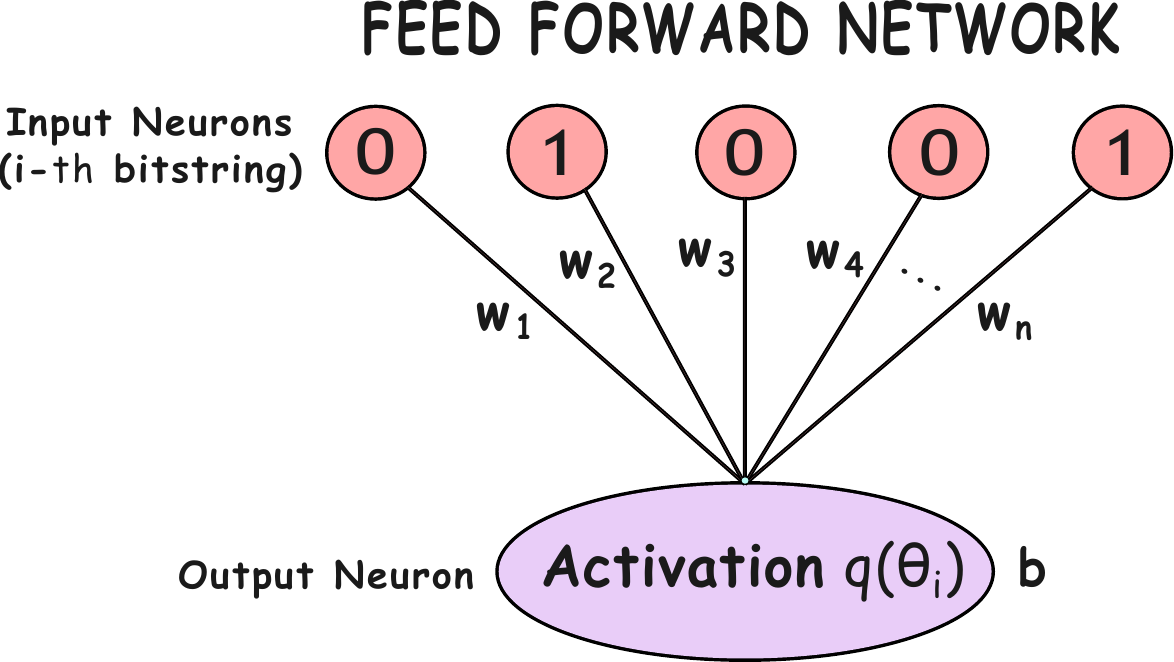}
\end{center}
\caption{\textbf{Visual depiction of a classical feed-forward neuron activation.} We show the
feed-forward structure of the neuron activation, which closely resembles a classical network containing trainable weights $w_{n}$
and bias $b$ on an individual example bitstring $i$. The activation function $q$ introduces non-linearity to the output neuron in the next layer.}
\label{fig: Feed Forward Neural Network}
\end{figure}

The goal of unsupervised generative modeling is to learn an unknown probability distribution $\mathcal{P}_{target}$ given a finite amount of samples such that it can produce new data from the same distribution \cite{bengio2017deep, gili2023generalization}. The QNBM, introduced by Gili et al. in 2022 \cite{GiliOG}, is a quantum generative analogue of the classical feed-forward neural network (FFNN - shown in Figure \ref{fig: Feed Forward Neural Network}) and carries over many of the latter's features.  
Like its classical counterpart, the QNBM is a multilayer network of units known as \textit{neurons} and is defined by its neuron structure, i.e. the number of neurons in each layer $(N_{in}, N_{hid_1}, N_{hid_2}, ..., N_{out})$. Each neuron within the model is assigned a qubit. 

The QNBM is comprised of an input register of qubits $\ket{x_{in}}$ (representing the first layer of the network), an output register $\ket{x_{out}}$ (final layer), optional hidden layers, as well as an ancilla qubit $\ket{0_{a}}$.  The ancilla (a component not present in a FFNN) is used for mapping activation functions from the input layer of neurons $\ket{x_{in}}$ to \textit{each output neuron} $\ket{\psi_{out}}$ in the next layer. A key difference between classical neural networks and the QNBM is the latter's ability to perform a non-linear activation function on a superposition of discrete bitstrings $\sum_{i}\ket{x_{in}}$. The non-linear activation function is applied through a sub-routine known as a Repeat-Until-Success (RUS) circuit. We also note that these non-linear activations distinguish the QNBM from other quantum generative models like the Quantum Circuit Born Machine (QCBM), which have previously been used to benchmark earlier versions of quantum hardware \cite{benny_qcbm}. QCBMs do not require mid-circuit measurement or classical control operations, which up until now, has made them easier to simulate and run on quantum hardware \cite{benny_qcbm, GiliQCBMGeneralize}. 

\SetKwComment{Comment}{/* }{ */}

\RestyleAlgo{ruled}
\begin{algorithm}[t!]
\caption{Repeat-Until-Success sub-routine}\label{alg:Repeat-Until-Success sub-routine}
\KwData{weights and a bias}
 Set success flag to false\; \\
\While{success flag is false}{
  Apply controlled rotation gates to ancilla\; \\
  Measure ancilla qubit\; \\
  \eIf{ancilla measurement is 0}{
    Set success flag to true\; \\
  }
  {
   Reset ancilla and output neuron qubit\; \Comment*[r]{Apply X and $R_{Y}$ gate}
  }
Move on to the next RUS sub-routine\;
}
\end{algorithm}

\begin{figure}[h!]
\begin{center}
\includegraphics[scale=0.85]{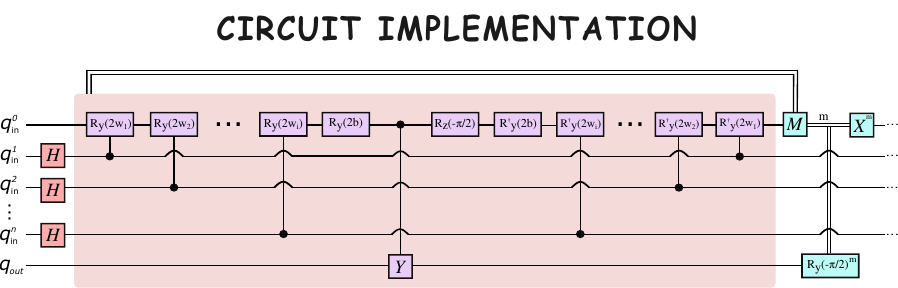}
\end{center}
\caption{\textbf{Quantum feed-forward neuron activation with non-linearity: Repeat-Until-Success Sub-Routine.} A non-linear activation is applied to a \textit{single} output neuron in a quantum generative model through a Repeat-Until-Success routine. We initialize all our input qubits (neurons) in a superposition. Then, we apply a series of gates parameterized by the weights $w_{n}$ and bias $b$. Finally, a measurement of the ancilla qubit is made. If the measurement is zero, the non-linear activation function $q$ has been applied to the output neuron. Otherwise, we must reset the ancilla and repeat the procedure.}
\label{fig: RUS Circuit}
\end{figure}

Like the FFNN, the QNBM has parameters - known as weights and biases - which, when tuned, determine its performance. A RUS circuit (Algorithm \ref{alg:Repeat-Until-Success sub-routine} and Figure \ref{fig: RUS Circuit}) performs the activation function at an output neuron through rotation gates parameterized by the neurons' weights (from the previous layer) and the output neuron bias. 
At the end of each RUS circuit, through mid-circuit
measurements of the ancilla, the following activation function is enacted on the respective output neuron:
\begin{equation} \label{eq: non-linear activation function}
q(\theta) = \arctan({\tan^2({2\theta})})
\end{equation}
The function parameter $\theta$ can be expressed as:
\begin{equation}
\label{eq: weights and biases sum}
\theta = w_{1}x_{1} + w_{2}x_{2} + \ldots + w_{n}x_{n} + b,
\end{equation}
where $w_{n}  \in (\text{-}1; 1)$ are the weights for \textit{n} neurons in the previous layer and $b \in (\text{-}1; 1)$ is the bias for a single output neuron.
\begin{remark}
    \textit{The non-linear activation function in Equation (\ref{eq: non-linear activation function}) contains a sigmoid shape, making it comparable to those typically used in \textit{classical neural networks} such as}
    \begin{equation}
        \sigma(x) = \frac{1}{(1 + e^{-x})}
    \label{eq: activation function}
    \end{equation}
\end{remark}
When a mid-circuit measurement is made on the ancilla qubit, if the state of the ancilla comes out to be $\ket{0_{a}}$ (which occurs with a probability of $p(\theta) > \frac{1}{2}$), the procedure is \textit{successful}. Therefore, the activation function on the output neuron has been performed. Otherwise, the activation function has not been applied, and we must (1) recover the pre-RUS circuit state with an X gate on the ancilla and $R_{Y}(\frac{-\pi}{2})$ applied to the output qubit and (2) repeat the RUS procedure until we receive a successful measurement. 
The final state of each RUS sub-routine with a successful activation can be described as:
\begin{equation}
\label{eq:successful activation state}
    \sum_{i} F_{i}\ket{x_{in}^{i}} \otimes \ket{0_{a}} \otimes R_{Y}(2q(\theta_{i}))\ket{\psi_{out}} 
\end{equation}
where $F_{i}$ refers to an amplitude deformation in the input state during the RUS mapping, and $\theta_{i}$ is the sum of weights and biases for each input bitstring.

\begin{remark}
\textit{We point out that the sum in Equation (\ref{eq:successful activation state}) comes from performing the RUS procedure on a superposition of bitstrings. We also note that there is no knowledge of how the amplitude of the input register $\ket{x_{in}}$ evolves under the RUS routine. Therefore, $F_{i}$ is unknown.}
\end{remark}

\begin{figure}[h!]
\begin{center}
\includegraphics[scale=1.5]{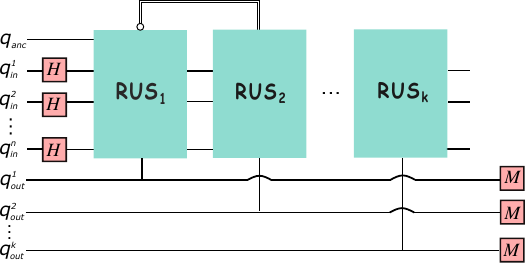}
\end{center}
\caption{\textbf{Quantum Neuron Born Machine.} Here we show the overall QNBM with multiple the Repeat-Until-Success (RUS) subroutines. An RUS routine is performed \textit{only} if the previous RUS routine was successful. Once all the routines are successful, we sample from the output neurons using the Born Rule to obtain our $P_{model}$.}
\label{fig: QNBM}
\end{figure}

Once all the output neurons $\ket{x_{out}}$ ($N_{out}$) have been connected to the input qubit register $\ket{x_{in}}$ ($N_{in}$) via their individual, successful RUS sub-routines, samples are drawn by measuring \textit{only} the output neurons according to the Born Rule (visually depicted in Figure \ref{fig: QNBM}). These samples are used to approximate the model’s encoded
probability distribution, i.e. $P_{model}$.
The QNBM is trained with a classical optimizer where $\theta$ (defined in Equation (\ref{eq: weights and biases sum})) is tuned to minimize the distance between $P_{target}$ and $P_{model}$ using a finite differences gradient estimator. This training scheme is very
similar to the ones used for other quantum generative models such as QCBMs \cite{QCBMDifferentiableLearning, GiliQCBMGeneralize, benny_qcbm}.

After training has concluded, samples are generated using the trained parameters for a separate evaluation of the model’s ability to learn the desired distribution. One can learn more about the emergence of the QNBM in Appendix \ref{sec:QNBM History}.  \\

\subsection{Stress Test Interpretation}
The intention behind \textit{stress testing} is to assess a system by pushing it beyond its specified thresholds to identify the load under which it breaks down \cite{StressTesting}. In software systems, stress tests can include an increase in concurrent virtual website users or a spike in client requests from a database. Stress testing can take on a slightly different interpretation when we consider stretching the load of a quantum system. In our case, the system to stress test is the Quantinuum H1 device, and we push the system by intentionally increasing the number of NISQ resources within the QNBM, an application-focused algorithm. In other words, using the QNBM model, we linearly scale the functional complexity by adding more RUS circuits (linear increase of input and output neurons) and then execute these various model sizes on the device. The total number of qubits required is equivalent to the total number of neurons, plus the ancilla. For each RUS circuit, there is an additional mid-circuit measurement of an ancilla, as well as an additional classical register to hold the value for a classically controlled operation. With each additional input-output neuron, there is an additional RUS circuit, where all other RUS circuits will have two additional parameterized two-qubit gates. For example, a $(2,0,3)$ contains $12$ parameterized two-qubit gates, $3$ fixed single qubit gates, and $3$ fixed two-qubit gates. Whereas, a $(3,0,4)$ contains $24$ parameterized two-qubit gates, $4$ fixed single qubit gates, and $4$ fixed two-qubit gates. Increasing the number of neurons also increases the number of measurement shot requirements as shown in Table \ref{shots}. Thus, one can see that increasing the neuron size of the QNBM places an incrementally larger quantum gate, classical register, and measurement (mid-circuit and final shots) load for stress testing. We then assess if the hardware is able to execute a model of a particular size, how well it performs based on the algorithm output, and the point at which the system reaches its maximum capacity. 

\section{Stress Test Quantitative Results}\label{quantitative}

Now that we have provided the stress test methodology with the QNBM, we implement it on Quantinuum's H1 machine. We provide details of this chosen hardware and the specific stress test that allows us to answer the following two questions: (1) How does the hardware perform on the QNBM and (2) How does the hardware's performance scale with the addition of more output neurons (RUS routines)? 

\subsection{Evaluation Details}

\paragraph{Hardware Description} In this work, we evaluate Quantinuum's Quantum Charged-Coupled Device (QCCD) architecture for the H-series hardware \cite{Pino2021}, which allows for high gate fidelities and scaling. Within the device, ions are arranged in linear chains called \textit{ionic crystals} confined on a surface trap. These crystals have the ability to be split, and their ions can be rearranged to execute entangling and single qubit gates on either a specific pair or individual ion, depending on the required operations. The ions can also be transported to other locations within the machine to minimize cross-talk when applying gates and measurements. 

\textit{So what happens under the hood?} When a quantum circuit is submitted from a user to the device, qubits are assigned to physical ions such that the number of transport operations is minimized. The circuit is executed by conducting a series of transport and gating procedures, suitably pairing and isolating ions for single-qubit and two-qubit gates, as well as measurements. Once all the operations are finished, the ions are restored to their original arrangement within the crystals, enabling the circuit to be repeated for the collection of measurement statistics. These results are then sent back to the user \cite{Pino2021}. 

Due to the hardware's ability to spatially isolate ionic crystals, mid-circuit measurements as well as gate operations can be performed on the target qubits with minimal risk of affecting the other ions, allowing for low cross-talk errors. The rearrangement and transport of ionic chains also allow for entangling operations on any pair of qubits, regardless of the spatial separation. This enables Quantinuum's machine to have all-to-all connectivity which reduces circuit depth and allows for high-fidelity computations.

For this work, we utilize the $Q-\#$ ($Q-\#$) \cite{Svore_2018} quantum software development kit to interface with the H1 hardware. We note that this is not the only way to interface with Quantinuum's hardware; we selected $Q-\#$ due to its prominence as a hardware-agnostic quantum programming language. We leave alternative QDKs, such as Pytket \cite{Sivarajah_2021}, for future work. 

Table \ref{Quantinuum_Specifications} below provides insight into the current error and fidelity rates of Quantinuum's H1 machine.

\begin{table}[h!]
\centering
 \begin{tabular}{||c r r r||} 
 \hline 
 Parameters & Min & Typ & Max \Tstrut\Bstrut\\ 
 \hline\hline 
 \rule{0pt}{4ex} Single-qubit gate infidelity & \num{1e-5} & \num{4e-5} & \num{3e-4} \\ [1ex]
 Two-qubit gate infidelity & \num{1.7e-3} & \num{2e-3} & \num{5e-3} \\ [1ex]
 State preparation and measurement (SPAM) error & \num{2e-3} & \num{3e-3} & \num{5e-3} \\ [1ex]
 Mid-circuit measurement and cross-talk error   & \num{5e-6} & \num{1e-5}  & \num{2e-4} \\ [1ex]
 \hline
 \end{tabular}
\caption{\textbf{Quantinuum H1$-$1 Specifications}} \label{Quantinuum_Specifications}
\end{table}

\begin{figure}[h]
\centering
\begin{subfigure}{.5\textwidth}
  \centering
  \includegraphics[width=1\linewidth]{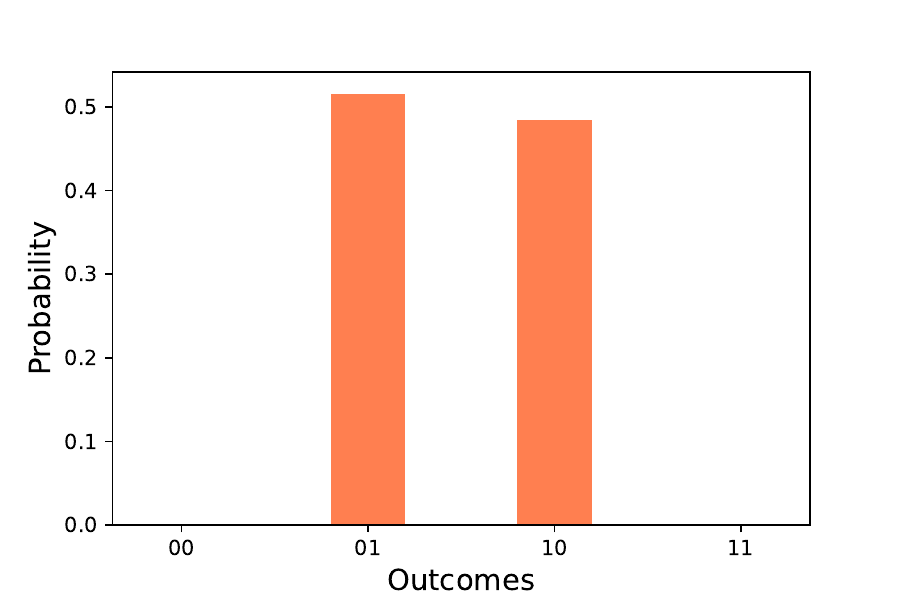}
  \caption{Target Distribution}
  \label{fig:target}
\end{subfigure}
\begin{subfigure}{0.5\textwidth}
  \centering
  \includegraphics[width=1\linewidth]{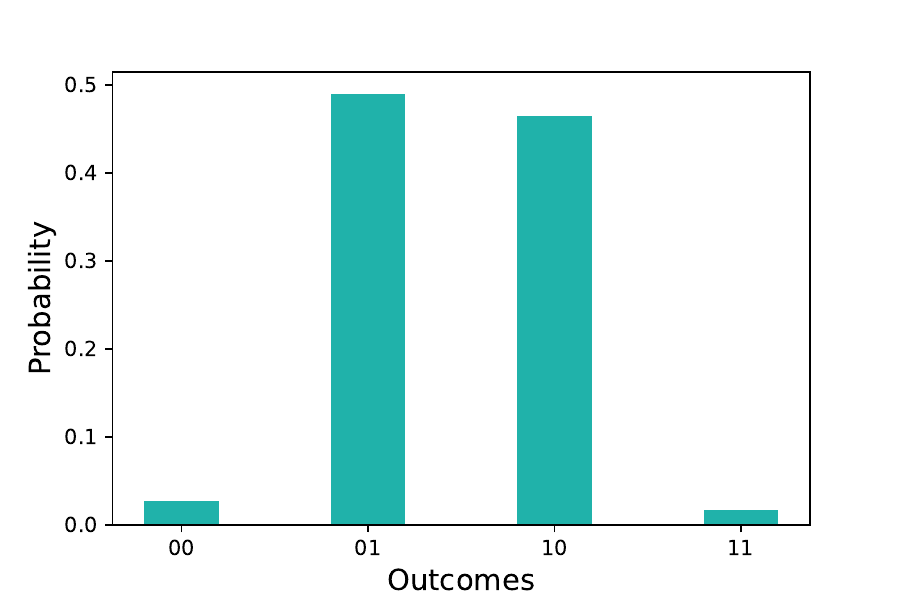}
  \caption{Emulator Execution}
  \label{fig:102emulator}
\end{subfigure}
\begin{subfigure}{0.5\textwidth}
  \centering
  \includegraphics[width=1\linewidth]{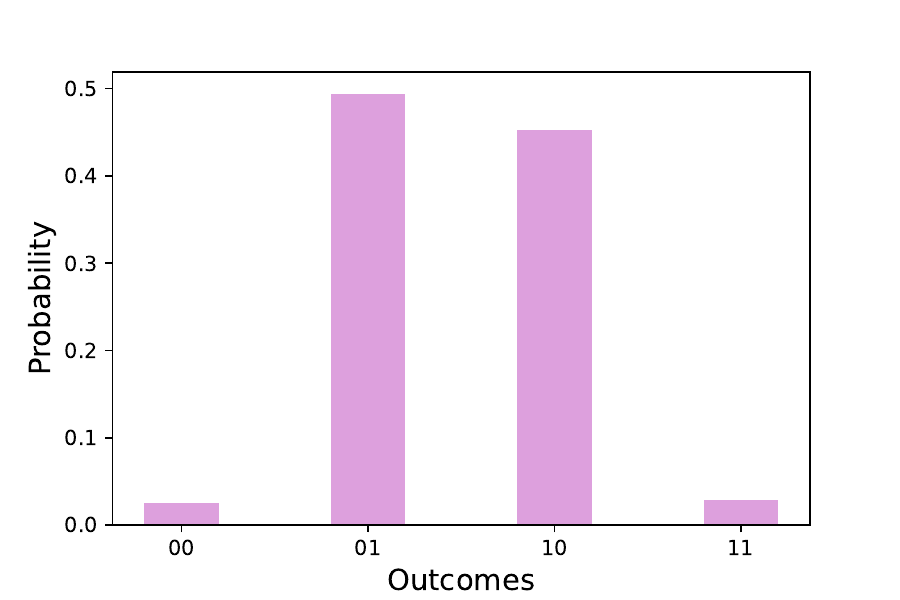}
  \caption{Hardware Execution}
  \label{fig:102HardwareRun}
\end{subfigure}
\caption{\textbf{Quantinuum’s H-1 hardware and emulator results for a 2 Output neuron QNBM algorithm.} We show the (a) target probability distribution and (b) emulator and (c) hardware results from running on Quantinuum's H-1 series device. We see that both the emulator and the hardware achieve high accuracy relative to the target, and that the emulator and hardware are near identical.  }
\label{fig:102Runs}
\end{figure}

\paragraph{Target Distribution}
\label{sec:picking-target-distribution}

To investigate the QNBM's performance on hardware, we first train the model to learn a target probability distribution $P_{target}$ of our choosing and then extract the trained parameters (weights and biases) that encode $P_{target}$. For our case, we choose a cardinality-constrained target distribution.
A \textit{cardinality-constrained distribution} is one that assigns the same probability $p$ to all bitstrings that contain a certain number $c$ of binary digits equivalent to 1. The probability $p$ is defined based on the set of bitstrings that fit this criteria, $\mathcal{S}$:
\begin{equation}
    p = \frac{1}{|S|}
\end{equation}
To satisfy the constraint of all probability distributions $\sum_{x} p(x) = 1$, all other bitstring probabilities will be zero. 

For our case, we choose the cardinality $c$ to be 1 for all target probability distributions given to our models. For instance, the target distribution for a (1,0,2) QNBM neuron structure where $c$ = 1 will be \{`00': 0.0, `01': 0.5, `10': 0.5, `11': 0.0\}.

\begin{remark}
    \textit{We note that, although a cardinality constrained distribution is a seemingly simple target distribution to choose, learning these distributions are a difficult and important task in combinatorial optimization \cite{gili2023generalization}}
\end{remark}

We utilize this target distribution to first obtain the optimal weights and biases for the model structure, using a high quality training procedure on the device simulator. The simulated training losses are demonstrated in Appendix Fig. \ref{fig: sim_training}; one can see high quality performance - i.e. low final KL values - for the parameters chosen to run on the hardware. To ensure that the hardware results are evaluated fairly, we compare the output of the hardware to a new target distribution $P'_{target}$, which is the model output distribution given from the simulated training. All target distributions that are compared to the hardware generated distributions in Sec. \ref{performance} are taken to be $P'_{target}$. 

\paragraph{Model Structure}
\label{sec:model-size-selection}

Given that we want to assess how the devices perform as we gradually scale the functional complexity of the algorithm - i.e. the size of the target distribution (number of output neurons) - we choose the following QNBM neuron structures: (1,0,2), (2,0,3), and (3,0,4). These specific structures contain no hidden layers, as networks without hidden layers have demonstrated the best training performance according to previous results in the literature \cite{GiliOG}.

We utilize 500 iterations to train all model structures, as these resource constraints were found to be sufficient for training QNBMs in Gili et al.\cite{GiliOG}. 

\paragraph{Training Metric}
\label{sec:training-metric}

Each model size is  trained on a noiseless simulator using the Kullback–Leibler (KL) Divergence \cite{info_theory_book} metric\footnote{Due to the size of models we consider, the KL Divergence metric is sufficient. For larger models, the KL Divergence scales poorly, and we encourage the use of other loss functions based on f-divergences \cite{f-train} and maximum mean discrepancies \cite{rudolph2023trainability}.} which quantifies the distance between the target probability distribution $P_{target}$ and the distribution being trained $P_{model}$. The KL metric is formally defined as
\begin{equation}
\label{eq: KL Divergence Metric}
    KL = \sum_{x} P_{target}(x)\log({\frac{P_{target}}{max(P_{model(x),\epsilon})}})
\end{equation}
where $\epsilon \approx 10^{-16}$ such that Equation (\ref{eq: KL Divergence Metric}) remains defined for $P_{model}(x)$ = 0.

The goal when training the QNBM model is to tune the weights and biases such that the distance between the distributions $P_{target}$ and $P_{model}$ converges to zero (KL $\approx$ 0). In other words, when the KL Divergence reaches a near-zero value, the model has learned and can express the target distribution.

\paragraph{Sampling Estimation}
\label{sec:resource-estimation-and-training}

Using post-selection, we obtain the following shot requirements. Given that we have $N_{out}$ output neurons, the model distribution (as well as the target) will be over $2^{N_{out}}$ possible bitstrings. Therefore, we need \textit{at least} $N_{shots}^{cc} = 2^{N_{out}}*K$ samples from the model to obtain the full probability distribution, where K is an arbitrary constant. 
If we have $N_{out}$ output neurons in our model, we have the same amount of RUS routines (given that there are no hidden layers). Each RUS circuit has $p_{N_{i} = success} = \frac{1}{2}$. Consequently, the probability that all $N_{out}$ RUS subroutines are successful is $p_{succ} = \frac{1}{2^{N_{out}}}$.

Therefore, if the expected amount of samples needed from the QNBM is  $2^{N_{out}}*K$ and the success probability of the entire model is $p_{succ} = \frac{1}{2^{N_{out}}}$, then the total amount of samples needed on the noiseless simulator using post-selection\footnote{In our work, post-selection involves executing all RUS circuits (regardless of ancilla measurement result). Then, we only select the trials where all ancilla measurements were successful to collect statistics.} is 
\begin{equation}
    N_{shots}^{ps} = (2^{N_{out}}*K)/p_{succ}
\end{equation}

In Table \ref{shots}, we show a comparison of the number of shots required using post-selection with the amount required utilizing mid-circuit measurement capabilities. 

\begin{table}[h!]
\centering
\begin{tabular}{lccc}
\toprule
& (1,0,2) &  (2,0,3) & (3,0,4) \\
\midrule
Classical Control     & 400 & 800 & 1600  \\ 
Post-Selection & 1600 & 6400 & 25600 \\     
\bottomrule
\end{tabular}
\caption{\textbf{Shot amounts required for each model size} depending on whether one uses post-selection ($N_{shots}^{ps}$) or classical control ($N_{shots}^{cc}$).}
\label{shots}
\end{table}

\begin{remark}
    \textit{We note that while we have used $N_{shots}^{cc} = 2^{N_{out}}*K$ ($K$=100) as our analytical formula for the number of shots (for classical control), we have not explored the trade-offs between the cost and performance as we tune $K$ to be smaller than 100.} 
\end{remark}

\begin{figure}[h!]
\begin{subfigure}{.5\textwidth}
  \centering
  \includegraphics[width=1\linewidth]{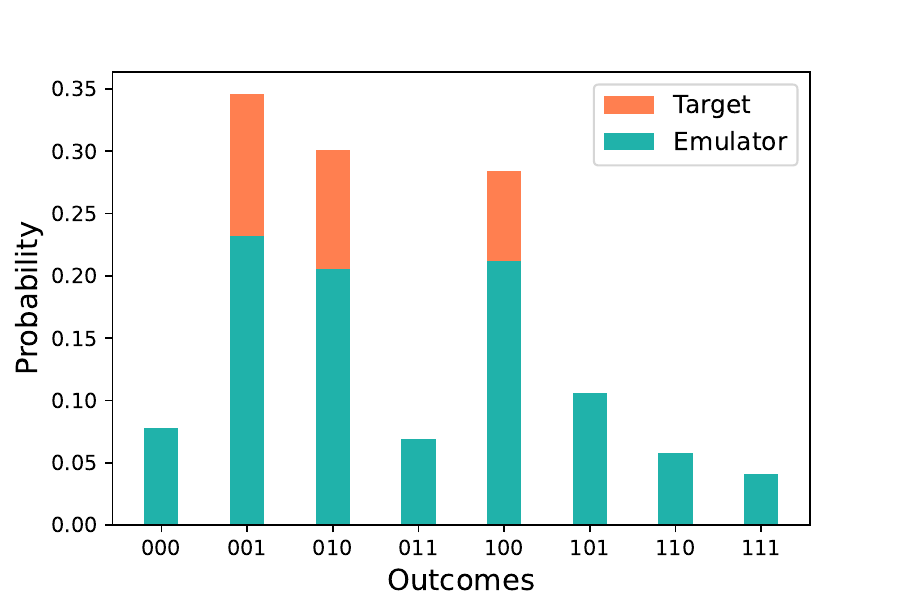}
  \caption{3 Output Neuron Emulator Execution}
  \label{fig:203HardwareRun}
\end{subfigure}
\begin{subfigure}{0.5\textwidth}
  \centering
  \includegraphics[width=1\linewidth]{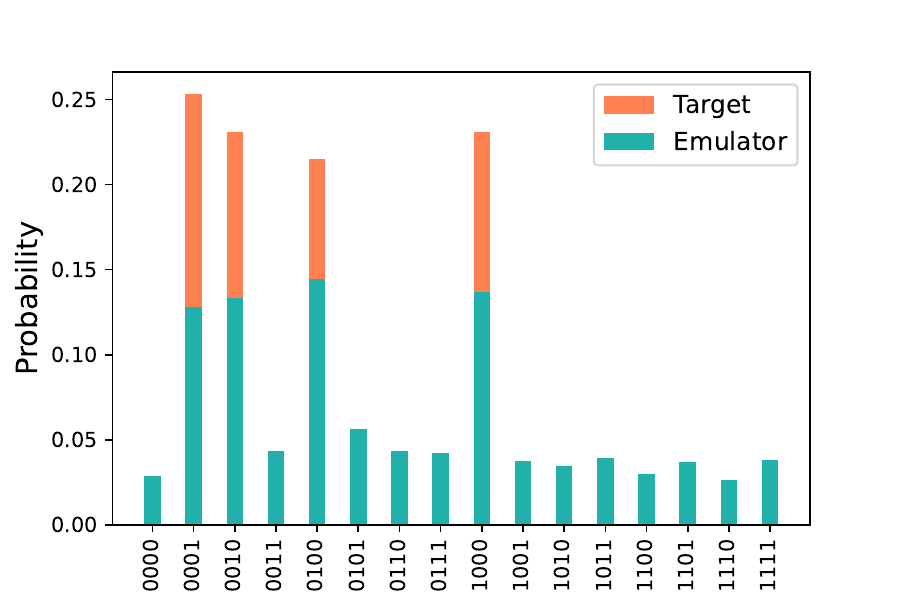}
  \caption{4 Output Neuron Emulator Execution}
  \label{fig:304HardwareRun}
\end{subfigure}
\caption{\textbf{Emulator results for three and four output neuron QNBM algorithms.} Due to the similar performance of the emulator and hardware as well as limited quantum resources, we find it sufficient to execute the larger-scale models on the emulator. We demonstrate the output probability distributions (green) relative to the target (orange) when scaling the number of output neurons. The performance only worsens from a $(1,0,2)$ as we scale the functional complexity of the algorithm, and still, we are able to see with decent clarity the model output bitstrings that were relevant to the problem over noise.  }
\label{fig:LargeScaleModelHardware}
\end{figure}

\subsection{Evaluation Questions}\label{performance}
\begin{figure}[h!]
\begin{center}
\includegraphics[scale=0.65]{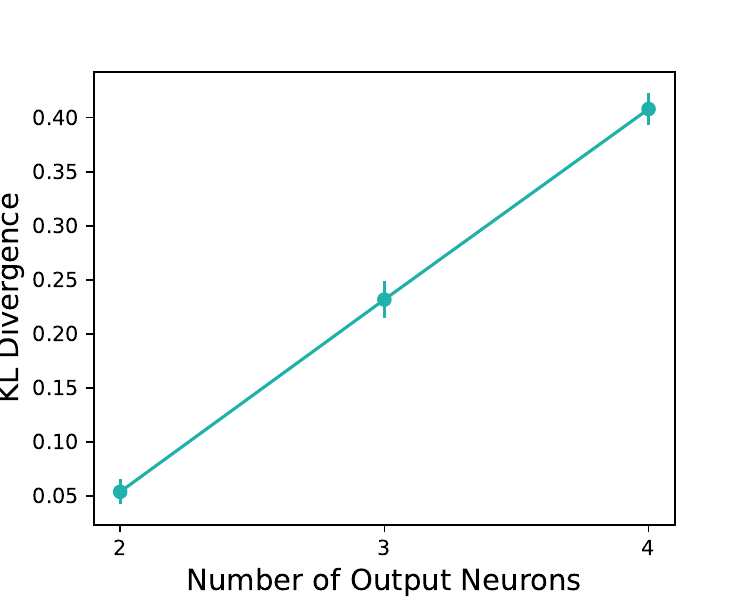}
\end{center}
\caption{\textbf{The relationship between the KL Divergence and increasing functional complexity of the algorithm (the number of neurons).} With three model sizes, we detect a linear trend in the data and see that the hardware's performance seems to moderately decrease according to the function $y = 0.178x + 0.053$ as we scale the number of mid-circuit measurement and classical control operations.}
\label{fig: KL Divergence Trend}
\end{figure}

\begin{center}
    \textit{1. How does the hardware perform when executing an algorithm of this complexity?}
\end{center}
Here we demonstrate the high-quality probability distribution from the execution of the (1,0,2) QNBM model with the trained parameters on Quantinuum's H1 QPU and the QPU emulator, comparing both results with the target distribution $P'_{target}$. As shown in Figure \ref{fig:102Runs}, the results from the hardware closely approximate the target distribution, reaching a KL value of 0.053, which indicates a strong overlap between distributions. This result highlights the exceptional performance of the algorithm on the hardware, specifically the novel classical control and mid-circuit measurement functionalities. Additionally, the results from the emulator and hardware have a massive overlap (KL = $0.0028$). Therefore, due to scarce quantum and classical resources, we found it sufficient to execute the larger-scale models (2,0,3) and (3,0,4) on the emulator. We ran 8 independent trials on the emulator and chose the best performing model to display; the small standard deviation bars are present in Figure \ref{fig: KL Divergence Trend}. As shown in Figure \ref{fig:LargeScaleModelHardware}, we can see that the generated probability distribution using classical control begins to diverge from the target distribution as the model size grows larger, and still, we see similar KL values for a $(2,0,3)$ and a $(3,0,4)$. For a $(2,0,3)$, we observe a KL value of $0.37$ and for a $(3,0,4)$, we observe a KL value of $0.39$. Even though these KL values are higher than a $(1,0,2)$, one can easily identify the bitstrings that the model was executed to output.   

\begin{center}
    \textit{2. How does the hardware perform when we scale the algorithm complexity \\ and add more RUS routines?}
\end{center}

Here, we show how the KL values of the target distribution $P'_{target}$ and the hardware/emulator output distribution scale as we increase the complexity of the algorithm via the number of output neurons. Again, we note that scaling the number of neurons directly translates to a linear increase in the number of classical control and mid-circuit measurement operations of the algorithm. Given Figure \ref{fig: KL Divergence Trend}, we can see the linear leap in the KL values when scaling the number of output neurons. With this small set of data, we observe a linear trend-line of $y = 0.178x + 0.053$. It is difficult to make any concrete claims on scaling here because we have such a small set of data to analyze; although we can conclude that, as the model grows larger and despite the low noise levels in the hardware, the algorithm's performance does suffer. However, even with the higher KL values for these slightly larger models, we can still distinguish the cardinality constraint on the distribution by comparing the probability weights of the different bitstrings. For instance, in Fig \ref{fig:304HardwareRun}, one can see that the probabilities for the bitstrings 0001, 0010, 0100, and 1000 are significantly higher than the rest, despite the divergence from the target distribution. In future work, when the hardware is resource-ready for larger scale models, we can obtain a more clear trend in the decrease of the loss with respect to the number of neurons. 

\section{Stress Test Qualitative Insights} \label{qualitative}
\label{sec:Challenges in Executing on Hardware}
We have demonstrated the quantitative results of the QNBM on the Quantinuum Ion-Trap H-Series Hardware, and observed probability distributions that produce a low KL Divergence value with respect to the target outcome. Despite the low noise levels and error rates within the hardware, we want to highlight that high-fidelity devices are not the only step needed for the full-scale realization of quantum computers. Here, we provide specific learning insights from our execution process with $Q-\#$ software that indicate the specific breaking points of the software and hardware, which call for further improvement.

\paragraph{Incompatibility between QDKs}
\label{sec: Incompatibility between QDKs}
To use classical control with Quantinuum's hardware, the creation of the quantum circuit as well as the state preparation and measurement (SPAM) component were ported from Python (Qiskit) to the $Q-\#$ language. 
On the surface, this may not seem like a massive issue. However, it is indeed non-trivial due to the differences between the Quantum Development Kits (QDKs). In classical devices, when creating algorithms, we normally do not think about whether our computers will support certain data structures available within a programming language (e.g., arrays, mutable variables, etc.). However, when using the H1 quantum device, this is indeed a limitation. In other words, many constructs that are available in the Python (Qiskit) framework were not compatible with the quantum hardware, such as indefinite loops. Therefore, if one wishes to utilize features such as classical control, porting between these QDKs requires a fair amount of manual labor and adjustment of the algorithm. 

\paragraph{Interfacing between Python and $Q-\#$} 
\label{sec: Interfacing between Python and Q-Sharp}
As shown in Algorithm \ref{alg:Repeat-Until-Success sub-routine}, the training of the QNBM contains a feedback loop. In other words, the parameters (weights and biases) are given to the model, which, when given the input, will output a probability distribution $P_{model}$. Then, the KL divergence between $P_{model}$ and $P_{target}$ is calculated. Based on the value of the metric, the parameters are tuned using the gradient. 

\begin{figure}[h!]
\begin{center}
\includegraphics[scale=0.45]{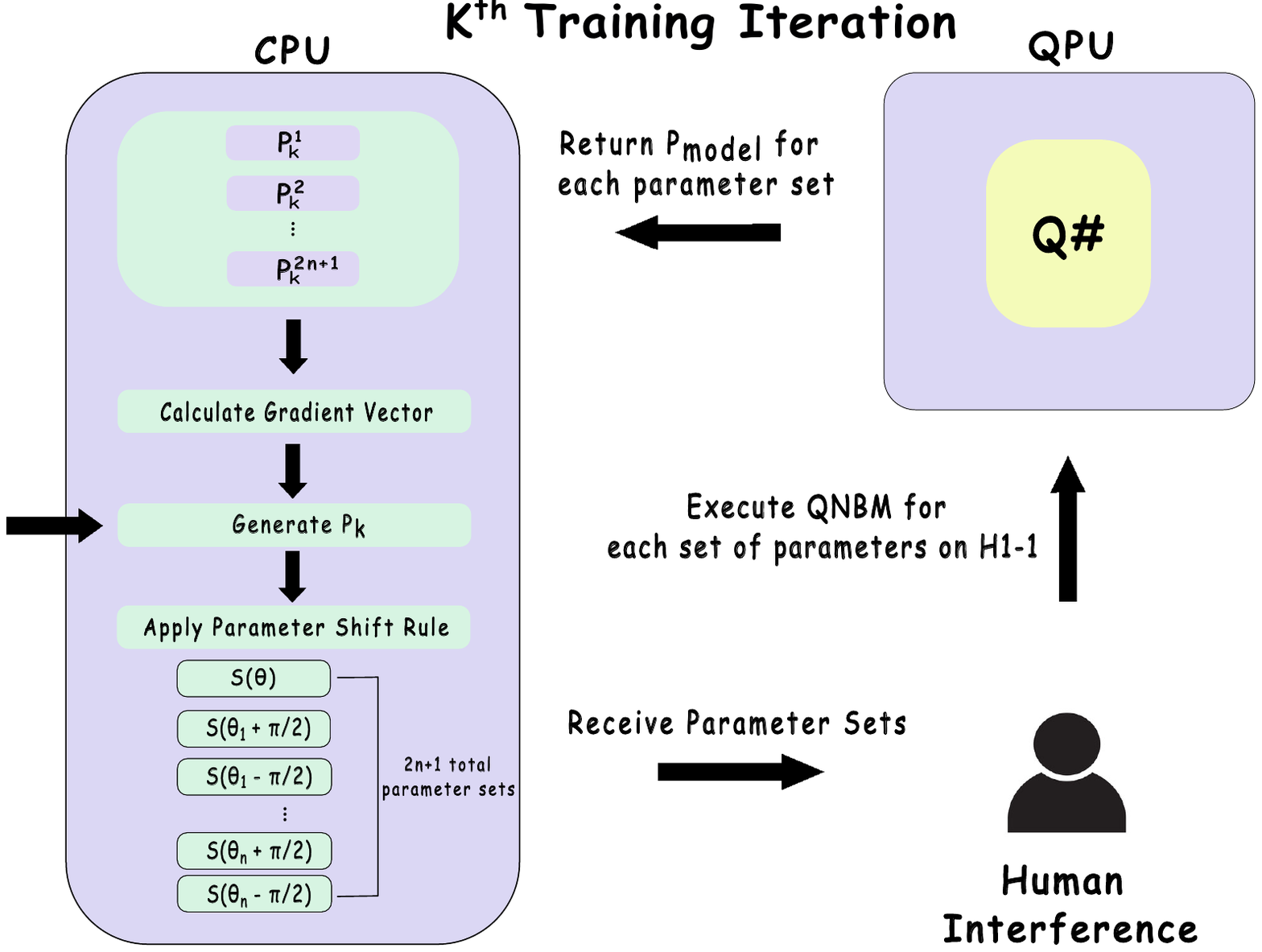}
\end{center}
\caption{\textbf{The training procedure on Quantinuum H1$-$1 Emulator.} To perform training on the H1 Emulator and work around interfacing issues between Python and $Q-\#$, a manual procedure needed to be done. For a single training iteration $k$ and given the generated probability distribution at $k$ ($P_{k}$), we: (a) apply the parameter shift rule for each parameter in $P_{k}$ such that we obtain  2n+1 new parameter sets, where $n$ is the number of params in $P_{k}$ (b) manually execute the QNBM using each parameter set on the Quantinuum QPU Emulator separately (c) using the output probability distributions for all 2n+1 parameter sets, calculate the gradient vector (d) generate the new set of probabilities $P_{k+1}$. Due to the extensive amount of iterations required to train a larger-scale QNBM model, the process quickly became infeasible to perform on the Emulator. Therefore, for this work, we choose to train the model on a noiseless simulator given by IBM and execute the model with the trained parameters on the H1 device. }
\label{fig: QNBM Training Procedure on H1-1}
\end{figure}

In our case, for this feedback loop, the QNBM quantum circuit was contained within the $Q-\#$ file, while the optimization was performed within Python. Due to the specialized functionality of $Q-\#$ , the $Q-\#$  code had to interface with the optimization written in Python. However, there were underlying issues within the files (that facilitated this communication) which prevented this interfacing. Additionally, $Q-\#$ functions did not support parameters sent from Python. Consequently, all optimization for training had to be done manually instead of through Python's built-in optimizers. Therefore, to train, the following process was formulated (Figure \ref{fig: QNBM Training Procedure on H1-1}):
\begin{itemize}
    \item the parameter set $p_{1}$ is hardcoded into the $Q-\#$  file, where $p_{1}$ is the initial set of weights and biases
    \begin{equation}
        p_{1} = \{{w_{1}, w_{2}, ...,w_{n}, b_{1}, b_{2},...,b_{k}}\}
    \end{equation}
    \item $Q-\#$  file is executed through the Azure Command Line tools
    \item From the $Q-\#$  file, the QNBM outputs a probability distribution $P_{model}$
    \item Post-process results 
    \item Take the gradient of the parameters $p_{1}$ using the parameter shift rule \cite{PhysRevA.99.032331}
    \item Obtain a new parameter set $p_{2}$
    \item Repeat process for $n$ iterations
\end{itemize} 

Given that there are a minimum of 500 iterations required to train larger models such as a (2,0,3) and a (3,0,4) \cite{GiliOG}, this process quickly became infeasible as it required a massive amount of human interference. 

\paragraph{Limited Classical Resources}
\label{sec: Limited Classical Resources}
Within a quantum device, there are quantum registers that define the qubits, and there are classical registers used for holding runtime values such as integers or supporting conditional branches. Thus, as the problem size of the model grew, the target machine H1 did not have enough classical registers to hold non-qubit values. Although counterintuitive, classical resources became a bottleneck for the quantum algorithm. 

In prior literature, it has been shown that the average amount of executions of an RUS circuit needed for success is bounded above by seven \cite{cao2017quantum}. However, due to the constraint of classical resources, we instead bound our Repeat-Until-Success algorithm above by six, giving the algorithm fewer branches and tries to succeed. Additionally, the resource constraint also placed a limit on the model sizes we could execute on the quantum hardware. Therefore, we found that executing a (3,0,4) model was the maximum size for the capacity of the current quantum device.

\paragraph{Financial Cost}
\label{sec: Cost}
Finally, executing on ion trap hardware is expensive and testing large-scale QNBM models beyond a (3,0,4) is simply too financially infeasible. The costs for each model size with the respective shot amount (Table \ref{shots}) are shown in Table \ref{shot_cost} as H-System Quantum Credits (HQCs).

\begin{table}[h!]
\centering
\begin{tabular}{lccc}
\toprule
& (1,0,2) &  (2,0,3) & (3,0,4) \\
\midrule
Classical Control     & 93 & 500 & 1570  \\ 
Post-Selection & 36 & 269 & 1822 \\     
\bottomrule
\end{tabular}
\caption{\textbf{Cost given in H-System Quantum Credits (HQCs)} to execute each model size depending on whether we perform post-selection or classical control. It is important to note how for smaller-scale models, post-selection seems more feasible and, as the model size grows, the benefit of classical control becomes apparent.}
\label{shot_cost}
\end{table} 

\section{Conclusion}
We have provided a quantitative and qualitative performance evaluation of Quantinuum's H-1 device using a novel stress test method with the QNBM algorithm. Based on the hardware's performance results, it is clear that there is remarkable small-scale capability with complex functionalities with decent scaling limitations. These results demonstrate excellent promise, and still we note that high-fidelity qubits, low error rates, and advanced functionality are not the only important factors for full-scale quantum computers. As shown in Sec. \ref{sec:Challenges in Executing on Hardware}, we face a plethora of challenges and additional manual labor when attempting to execute any algorithm on hardware. It is not our intention to discourage individuals from executing algorithms on the devices. Rather, it is to raise awareness that, alongside high-quality quantum hardware with more advanced operations, every layer within the quantum computing stack must interface seamlessly for full-scale realization. Lastly, we encourage researchers to use the stress test interpretation presented in this work to evaluate other advanced NISQ devices as they improve and scale.

\begin{figure}[h]
\begin{center}
\includegraphics[scale=0.45]{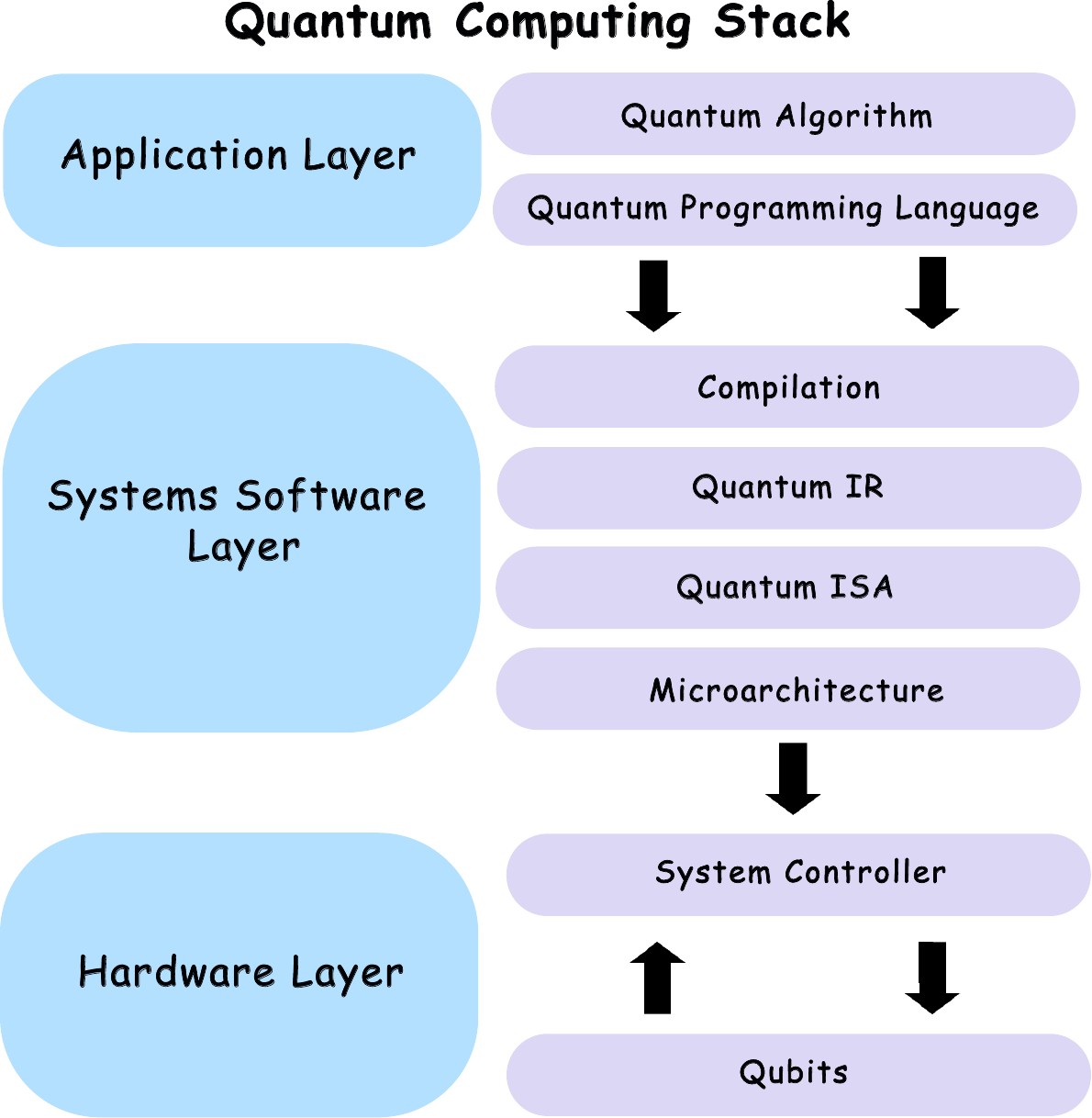}
\end{center}
\label{fig: QPU Stack}
\end{figure}

\section{Data Availability}
All source code used to collect data contained in this article is located in the following Github repository: https://github.com/kaitlinmgili/qnbm.

\section{Acknowledgements}
AUS would like to acknowledge Agnostiq Inc, Aspirant Ventures, Atmos Ventures, the CUbit Quantum Initiative, EeroQ, Multiverse Computing, the Quantum Computing Report, The Center for Computation and Technology at Louisiana State University, and Rigetti for financial support. Additionally, many thanks to Denise Ruffner and the Women in Quantum donors for providing support for AUS. We would like to thank Dhrumil Patel, Jamie Leppard, Mario Gely, Mykolas Sveistrys, and Rohan Kumar for insightful discussions. Special thanks to the team at Microsoft Azure for working with us to implement the algorithm on the ion trap H-Series hardware. Additionally, we thank the Womanium Quantum summer program for providing insight into the different hardware platforms. 

\section{Appendix}

\subsection{Simulator Training}

Here, we show the training loss curve that provides the optimal weights and biases to utilize for our model on the H1 hardware and emulator devices. 

\begin{figure}[h!]
\begin{center}
\includegraphics[scale=0.4]{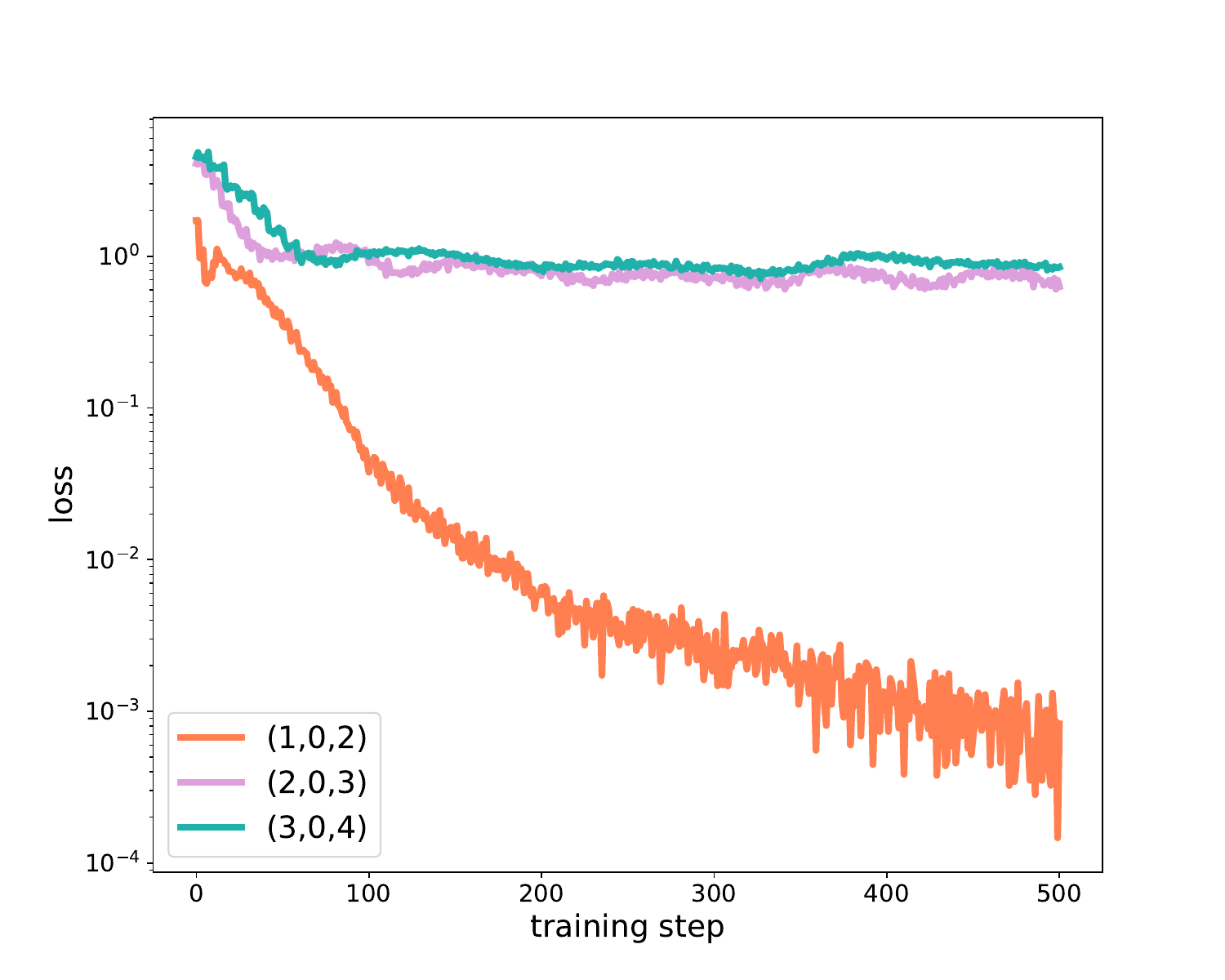}
\end{center}
\caption{\textbf{The KL training curves for all model sizes on the statevector simulator.} Here, we show the convergence of a $(1,0,2)$ (orange), a $(2,0,3)$ (pink), and a $(3,0,4)$ (green) QNBM algorithm. The parameters utilized for the hardware circuit are taken from the last iteration step. Consequently, the output probability distribution at the final training step becomes the target distribution for stress testing the H1 hardware/emulator. }
\label{fig: sim_training}
\end{figure}

\subsection{Emergence of the QNBM}
\label{sec:QNBM History}
In recent years, the field of quantum machine learning (QML) has emerged to explore the intersection between quantum computation and classical machine learning algorithms. Specifically, QML aims to (1) study how we can leverage techniques in machine learning to unearth new ideas in quantum information processing and (2) understand how quantum machines can potentially exhibit advantageous behavior over their classical counterparts in data-driven tasks.
For the latter purpose of QML, one of the most encouraging research directions for achieving practical quantum advantage is \textit{unsupervised generative modeling}. \cite{benny_qcbm, gili2023generalization, QCBMDifferentiableLearning}

The goal of unsupervised generative modeling is for a machine to learn an unknown underlying probability distribution from an unlabeled training data set such that it can produce new data samples from the same distribution. Nowadays, we can see large-scale classical generative models being used for challenging tasks such as recommendation systems \cite{cheng2016wide}, drug discovery \cite{nica2022evaluating}, and image generation \cite{styleGAN}. Ultimately, developing powerful generative models is imperative to realizing advanced artificial intelligence applications, including image classification and recommendation systems \cite{gen_AI_apps}. 

\subparagraph{Generative Modeling using Quantum Computers}
\label{sec:Generative-Modeling-using-Quantum-Computers}

\textit{How does quantum computing fit into unsupervised generative modeling?} Recall that a quantum logical circuit consists of several quantum bits (qubits) that, when measured in the computational basis, each give a probabilistic outcome of 0 or 1. Repeating the process of circuit execution and measurement of the quantum bits several times (known as sampling) can create a probability distribution over a set of finite bitstrings. For instance, for a circuit with $N$ qubits, sampling from this circuit would create a probability distribution over $2^{N}$ possible outcomes. With the addition of parameterized quantum logic gates within the circuit, the resulting probability distribution can actually be \textit{tuned} by changing the parameters. Therefore, since these circuits intrinsically produce probability distributions through measurements, utilizing them for the purpose of generative modeling is a natural idea.  

\textit{How do we arrange these parameterized gates within the circuit?} Numerous architectures for these quantum generative models have been proposed \cite{Zhu_2019, cherrat2022quantum, Zoufal_2019, Romero_2017} however, the optimal structure for the NISQ and fault-tolerant eras has yet to be found. 

A widely known quantum generative model that uses parameterized quantum circuits is the Quantum Circuit Born Machine (QCBM)\cite{QCBMDifferentiableLearning,GiliQCBMGeneralize}. However, this model outputs probability distributions through projective measurements of quantum states that have evolved \textit{linearly}. Various classical machine learning models today employ non-linearity to capture complex relationships between data. 

The QNBM is the first quantum generative model to apply a non-linear activation function to a register of qubits. Previous literature has explored various aspects of the QNBM, including the model's learning capabilities, classical non-simulatabilty, how its performance compares to classical models (such as Restricted Boltzmann Machines), and its superior performance over the QCBM \cite{gili2023generative}. 
\\

\printbibliography

\end{document}